%% file: main.tex
\documentclass{vgtc}                  
\ifpdf
  \pdfoutput=1\relax                   
  \usepackage{graphicx}                
  \DeclareGraphicsExtensions{.pdf,.png,.jpg,.jpeg} 
\else
  \usepackage{graphicx}                
  \DeclareGraphicsExtensions{.eps}     
\fi%

\graphicspath{{figures/}{pictures/}{images/}{./}} 

\usepackage{microtype}                 
\PassOptionsToPackage{warn}{textcomp}  
\usepackage{textcomp}                  
\usepackage{mathptmx}                  
\usepackage{times}                     
\usepackage{cite}                      
\usepackage{tabu}                      
\usepackage{tabularx}
\usepackage{changepage}
\usepackage{booktabs}                  

\usepackage{enumitem}
\setlist[itemize]{noitemsep, topsep=0pt}
\usepackage{wrapfig}
\usepackage{caption}
\usepackage{titlecaps}
\usepackage{dblfloatfix}
\usepackage{amsmath}
\usepackage{amssymb}
\usepackage{newtxmath}
\usepackage{fontawesome}
\usepackage{placeins}
\usepackage{color}
\usepackage{physunits}

\usepackage{etoolbox}
\AtBeginEnvironment{quote}{\par\singlespacing}

\definecolor{cb_orange}{rgb}{1.0,0.51,0.0}
\definecolor{cb_blue}{rgb}{0.22,0.49,0.72}
\definecolor{cb_green}{rgb}{0.3,0.67,0.29}
\definecolor{cb_red}{rgb}{0.89,0.1,0.11}
\definecolor{cb_purple}{rgb}{0.6, 0.31, 0.64}
\definecolor{cb_brown}{rgb}{0.6, 0.4, 0.2}
\definecolor{cb_crimson}{rgb}{0.86, 0.08, 0.24}

\def\genlens{\texttt{GenLens}}

\onlineid{1103}

\vgtccategory{Research}
\vgtcpapertype{Application/Design Study}

\title{GenLens: A Systematic Evaluation of Visual GenAI Model Outputs}

\author{Tica Lin\thanks{e-mail: mlin@g.harvard.edu}\\ %
        \parbox{1.3in}{\vspace{1mm}\scriptsize \centering Harvard University \\ Adobe Research}%
\and Hanspeter Pfister\thanks{e-mail: pfister@g.harvard.edu}\\ %
     \scriptsize Harvard University %
\and Jui-Hsien Wang\thanks{e-mail: juiwang@adobe.com}\\ %
     \scriptsize Adobe Research}

\input{docs/0_abstract}

\CCScatlist{ 
\CCScatTwelve{Human-centered computing}{Interaction design}{Interaction design process and methods}{User interface design};
\CCScatTwelve{Human-centered computing}{Visualization}{Visualization application domains}{Visual analytics}
}

\teaser{
 \includegraphics[width=1.1\linewidth]{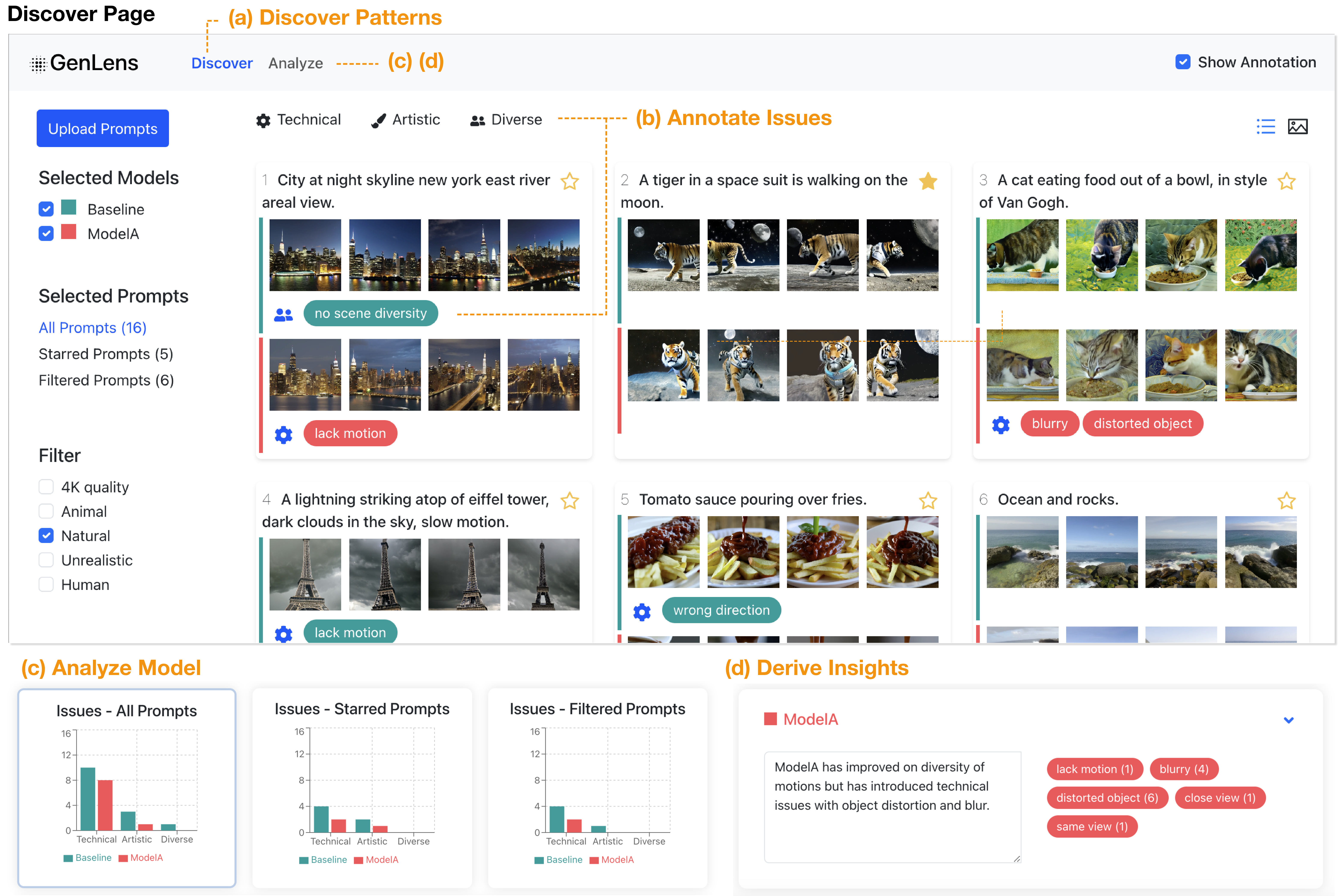}
 \caption{
 \genlens{} is an interactive web application designed to facilitate the annotation and analysis of visual outputs from GenAI models. This tool supports a comprehensive evaluation pipeline from (a) discovering patterns, (b) annotating issues, (c) analyzing model performance with aggregated results, and (d) deriving insights with quantifiable evidence for model training enhancement.
 \textit{Discover Page} interface supports (a), while \textit{Annotation Mode} (Fig.~\ref{fig:annotation}) supports (b) and \textit{Analyze Page} (Fig.~\ref{fig:analyze})  supports (c) and (d).
 }
    \label{fig:teaser}
}

\vgtcinsertpkg

\begin{document}


\input{docs/1_intro}
\input{docs/2_related_work.tex}
\input{docs/3_formative_study}

\input{docs/4_design.tex}
\input{docs/5_discussion.tex}
\input{docs/6_conclusion.tex}

\input{docs/acknowledgement}

\bibliographystyle{abbrv-doi-hyperref}
\bibliography{references}
\end{document}

%% file: docs/0_abstract.tex
\abstract{
The rapid development of generative AI (GenAI) models in computer vision necessitates effective evaluation methods to ensure their quality and fairness. Existing tools primarily focus on dataset quality assurance and model explainability, leaving a significant gap in GenAI output evaluation during model development. Current practices often depend on developers' subjective visual assessments, which may lack scalability and generalizability.
This paper bridges this gap by conducting a formative study with GenAI model developers in an industrial setting. Our findings led to the development of \genlens{}, a visual analytic interface designed for the systematic evaluation of GenAI model outputs during the early stages of model development. \genlens{} offers a quantifiable approach for overviewing and annotating failure cases, customizing issue tags and classifications, and aggregating annotations from multiple users to enhance collaboration. 
A user study with model developers reveals that \genlens{} effectively enhances their workflow, evidenced by high satisfaction rates and a strong intent to integrate it into their practices. This research underscores the importance of robust early-stage evaluation tools in GenAI development, contributing to the advancement of fair and high-quality GenAI models.
} 

%% file: docs/1_intro.tex

\firstsection{Introduction}
\maketitle
The advent of generative AI (GenAI) models, particularly in the domain of computer vision, marks a significant milestone in the evolution of artificial intelligence. GenAI models, using advanced neural network architectures, can learn from vast datasets to create new, original content that mimics the input data characteristics. 
These models have shown remarkable capability in generating realistic images, videos, and other types of visual content, pushing the boundaries of creativity and automation. However, the rapid pace of development in this area brings forth a critical challenge: the need for robust and effective evaluation methods to ensure the quality and fairness of these models.

Traditionally, visualization tools in AI have focused on ensuring dataset quality and facilitating model explainability~\cite{yuan2021survey,la2023state}.
While these explainable AI tools are vital, they predominantly address issues before model training or at later stages of model development, leaving a critical gap in the early stages of GenAI output evaluation. 
In the early stages of GenAI output evaluation, developers make critical design decisions to correct potential model design flaws, data handling issues, and algorithmic biases. 
This phase is critical for guiding the development process amidst intense competition and expensive data training.
In practice, the evaluation of these early-stage outputs often heavily relies on developers' subjective visual assessments. Though this approach is intuitive and allows fast iteration, it can lead to the oversight of crucial biases and potentially result in inefficient use of computational resources in later stages of the development cycle.

Recognizing this gap, our research aims to develop a systematic approach for the early evaluation of GenAI models. This paper presents the outcomes of a comprehensive formative study conducted with GenAI model developers working in industrial image and video generative models. Two gaps were highlighted, including effective ways to identify failure cases and verify the improvement across prompts and models.

Drawing on these findings, we developed \genlens{}, a collaborative visual analytic tool tailored for the evaluation of visual outputs from GenAI models. It features two key functions to support model output evaluation: discovering patterns and analyzing issues. We developed \genlens{} through an iterative design process and incorporated feedback from model developers to ensure a user-centered design. The final prototype was evaluated in a controlled user study with GenAI model developers using real model output data. The evaluation results demonstrate that \genlens{} effectively enhances model developers' workflows by providing an intuitive interface for annotating and comparing failure cases, along with the capability to share these findings. We later incorporated user feedback and deployed the tool for internal use, which will allow for more comprehensive feedback to be collected for future work. 

This research contributes to the field of GenAI and visualization by highlighting the importance of a systematic evaluation tool. We contribute to a first-hand understanding of the workflow and challenges faced by GenAI model developers
and provide a visual analytic solution, \genlens{}, to steer the model development process based on real user needs. 
Our results suggest that the \genlens{}  approach supports an effective model evaluation workflow and encourages collaboration for better insight verification. The paper concludes by discussing the broader implications of our findings on early-stage model evaluation and the importance of a human-centered approach in GenAI development.

%% file: docs/2_related_work.tex
\section{Related Work}
\textbf{Generative AI.} Recent advancements in AI have been heavily centered around generative AI (GenAI). This subset of AI creates new content based on learned patterns and data inputs, ranging from textual to visual outputs. Leveraging transformer neural network~\cite{vaswani2017attention}, GenAI language foundation models like GPT demonstrate high capability in complex natural language processing tasks like summarizing and reasoning. The scope of GenAI has further extended to the realms of image and video generation, showing great abilities in synthesizing visual data. Image foundation models like DALL-E generate detailed images from text and various multi-modal tasks~\cite{hertz2022prompt, brooks2023instructpix2pix, zhang2023adding}. More recently, GenAI's expansion into video creation is rapidly evolving~\cite{guo2023animatediff,chen2023videocrafter1}, targeting producing clips from text or editing existing videos.
While this offers immense potential, it also poses challenges in ethics and fairness, highlighting the need for rigorous evaluation of AI-generated content throughout and beyond the model development process.

\textbf{Visual analytics for machine learning.} With the rapid growth of AI, this research domain employs visualization techniques to enhance the explainability, trustworthiness, and reliability of machine learning. Primary goals focus on supporting the analysis and understanding of different components throughout the training process, including before, during, and after model building~\cite{yuan2021survey}. Much work  is dedicated to explaining model structures and interpreting outputs to understand model behaviors~\cite{wang2019deepvid,yuan2022context}.

In the realm of advanced deep learning networks and GenAI models, due to the complex model structures and generative nature of the outcomes, ensuring data quality and properly evaluating model outputs have become instrumental to steering model improvement~\cite{wang2023visual}.
In particular, visualization techniques are crucial for managing large-scale visual data~\cite{chen2020oodanalyzer}. Xie et al.~\cite{xie2018semantic} proposed a semantic-based approach for visualizing images and their embeddings with interactive navigation, which supports the exploration of large image collections.
To further optimize the semantic layout of large visual datasets, Bertucci et al.~\cite{bertucci2022dendromap} proposed a novel dendrogram treemap technique to visualize image datasets based on hierarchical clustering algorithms.
To analyze the model outputs, 
prior works targeted comparing performance between models and instances~\cite{li2018embeddingvis,cabrera2019fairvis}. Chen et al.~\cite{chen2017qsanglyzer} proposed QSAnglyzer to visualize model performance across different categories, enabling efficient model comparisons. 

Building upon previous work in visualization for machine learning, we have identified a significant gap in the early-stage evaluation of visual outputs from GenAI models. This work aims to address challenges faced by developers in effectively assessing model performance during model development.

%% file: docs/3_formative_study.tex

\section{Formative Study with GenAI Model Developers}
To better understand the current development workflow and challenges among GenAI model developers, we conducted a series of interviews with six researchers and engineers within our organization. Five of them hold PhD degrees in computer science.
They have worked on advanced computer vision model development with 3 to 10 years of experience, particularly in training GenAI models on visual data such as text-to-image, content-aware inpainting, or text-to-video models. 

Each interview elicited developers' experiences from four perspectives, including 1) overall model training goal and process, 2) data type involved in training and analysis, 3) tasks they performed to evaluate model output, and 4) current tools and gaps. Lastly, we asked interviewees for visualization features that could enhance their workflow. 

We summarize the workflow and gaps in visualization for GenAI model evaluation based on thematic analysis of the interview results.

\textbf{GenAI model development workflow.}
While the specific data types differ, these large-scale GenAI models workflow involves similar steps, including first collecting and training on a large dataset, testing the model on desired downstream tasks with examples (e.g., a prompt-to-image result pair), followed by analyzing failure cases to generate insights towards improving the model, such as the decision to resample the data.
Although most parts of model development only involve command-line interfaces, we found two specific areas requiring heavy use of visualizations, including curating datasets and analyzing model outputs.
 
Due to the large dataset, often at a million to billion number scale, strategic sampling of images and videos for visual inspection is required to ensure data quality and align data with the targeted use cases. Prior work has addressed visualizing large image collections to analyze embedding and hierarchy~\cite{xie2018semantic, bertucci2022dendromap}. Specific visualization tools for data management were used to support the model development. 

However, gaps were found in effective ways to support evaluating outputs. The current evaluation of the model performance was purely based on expert intuition and manual inspection. To identify issues, developers iteratively query and analyze examples based on previous error cases. 
Due to the lack of effective visualization tools, they often had to manually build visualization ad-hoc to search for specific examples and verify their hypothesis, leading to difficulties in sharing and verifying their insights across model iteration and collaboration.

\begin{figure}[t]
  \centering
  \includegraphics[width=\linewidth]{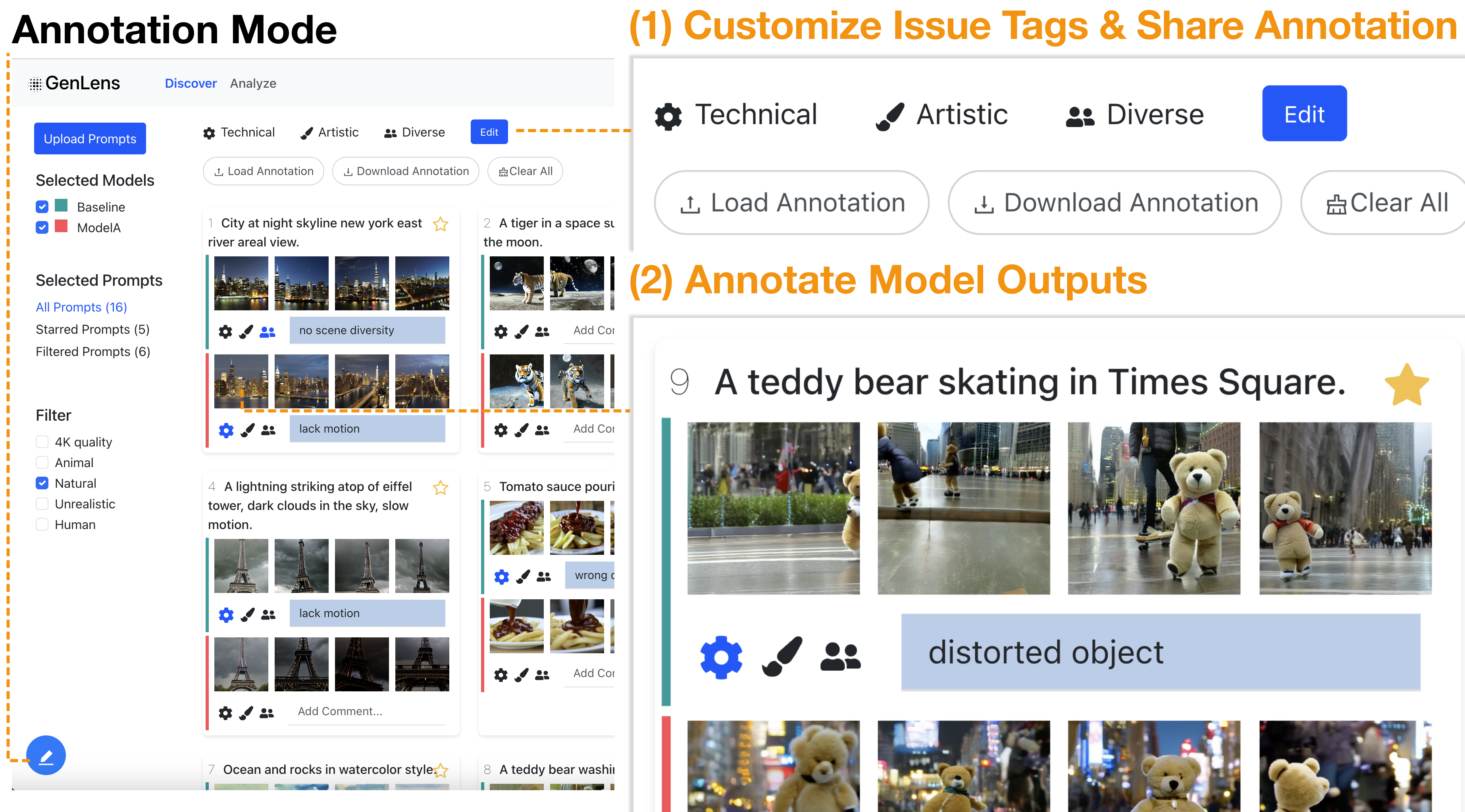}
  \caption{\textit{Annotation Mode} interface. Users can (1) specify their own issue tags and (2) annotate each output with icons and comments.}
  \label{fig:annotation}
\end{figure}

\textbf{Gaps in evaluating model outputs.}
Specifically, to support effective model output evaluation, four common goals need to be met. A developer needs to: G1) \textit{Discover patterns} by overviewing the model outputs. This helps developers obtain high-level impressions of the overall model performance. 
G2) \textit{Identify issues} by contrasting success and failure cases. This allows developers to examine specific problems with each output. G3) \textit{Analyze performance} by summarizing identified issues across examples. This step is crucial for developers to make observations at the model level. G4) \textit{Derive insights} by aggregating analysis from multiple developers and models. This supports more transparent and collaborative insight generation and verification. 

Based on the findings,
\genlens{} was designed to satisfy these critical goals during the model evaluation workflow (Sec.~\ref{sec:workflow}).
We further categorized three key visualization tasks based on the pain points faced by developers from our interviews, which inform the interface design of \genlens{} in Sec.~\ref{sec:interface}.

\textbf{T1) Explore and compare the outputs.} With diverse examples from a batch of model outputs, developers need to narrow down to focus on specific categories of interest, such as examples involving humans. Further, comparing the outputs with a baseline is important to allow grounded evaluation. While data are available, an effective interface is lacking to navigate the model outputs effectively.

\textbf{T2) Annotate issues across examples.} 
It is important to specify the type of issues in the failure cases to inform aggregated insights for improvement, such as technical or ethical issues. However, with limited resources, current developers only subjectively examine the results and take high-level notes on overall performance. A user-friendly and scalable solution for efficient annotation is thus required.

\textbf{T3) Quantify model performance.} In addition to evaluating the outputs at the instance level, developers need to summarize the overall performance on the model level.
Currently, developers observe insights purely based on subjective impressions. Without an explicit and quantifiable demonstration of identified issues, it is challenging to verify insights and track improvements across models. There is a need for better support in quantifying and sharing model evaluation results.

Overall, developers sought a more intuitive way to link and analyze failure cases to effectively evaluate GenAI model performance.

%% file: docs/4_design.tex
\section{GenLens} 
\label{sec:genlens}
Based on the findings from the formative study, we developed \genlens{}, a visual analytic interface for systematic GenAI model output evaluations. 
To address the current gaps, 
\genlens{} supports a quantifiable and scalable evaluation workflow from discovering patterns, specifying issues, analyzing performance to deriving insights. 

We first described a developer's evaluating workflow using \genlens{}, followed by the interface design of \genlens{}.

\begin{figure}[t]
  \centering
  \includegraphics[width=\linewidth]{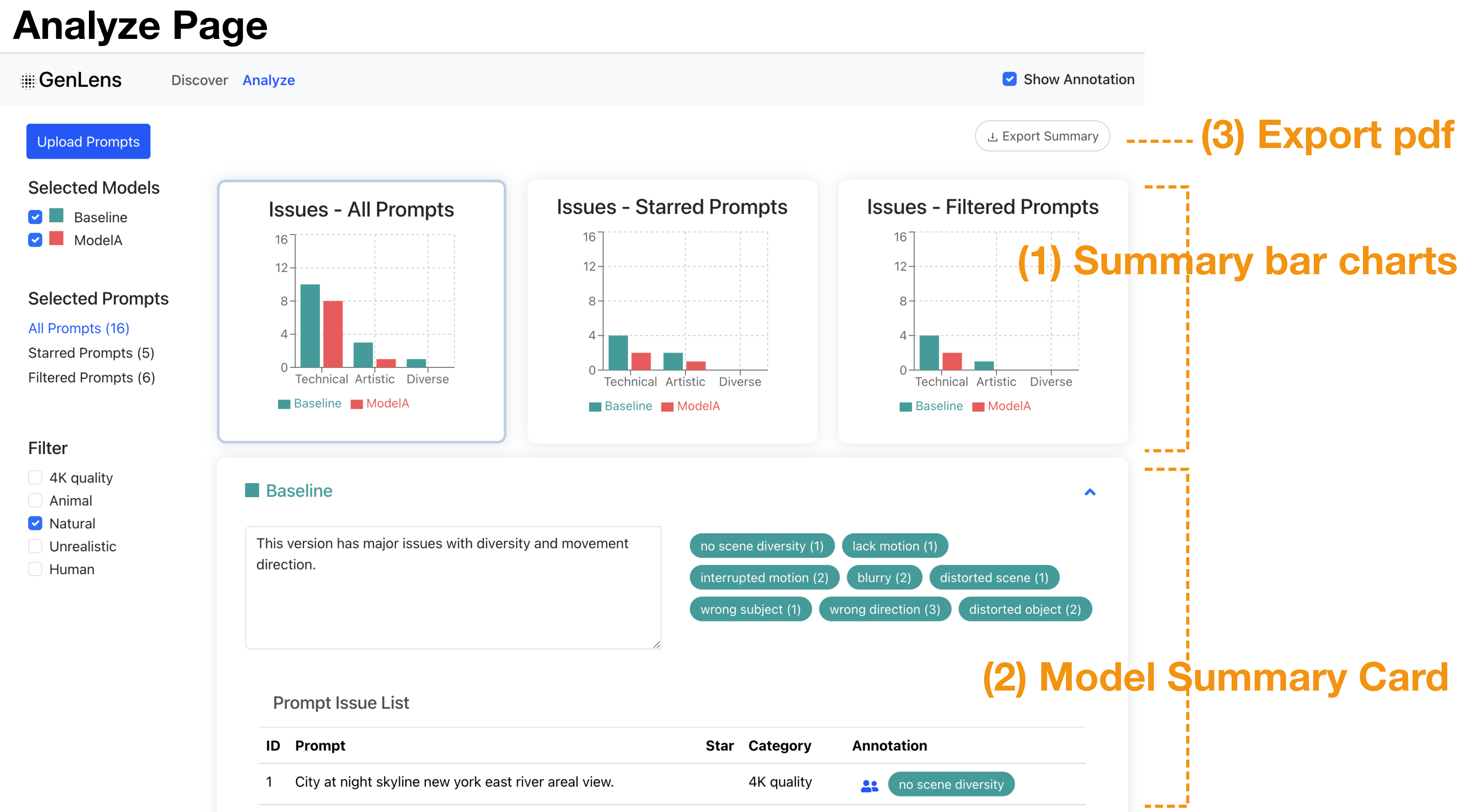}
  \caption{\textit{Analyze Page} interface. Users can (1) analyze model issues with summary bar charts and (2) derive insights towards model performance based on aggregated annotations in the model summary cards. (3) A summary report can be exported as pdf.}
  \label{fig:analyze}
\end{figure}

\subsection{Evaluation Workflow}
\label{sec:workflow}
Ben is a model developer working on training a vision GenAI model targeting text-to-video tasks.
After running a new model version on a predefined prompt set that was used across developers to evaluate model improvement over iterations, Ben uses \genlens{} to evaluate the video examples from model outputs. Fig.~\ref{fig:teaser} shows four key goals \genlens{} supports in Ben's workflow.

\textbf{G1. Discover Patterns (Fig.~\ref{fig:teaser}a).} On the \textit{Discover Page},
he first overviews all the results and compares the new model outputs ($ModelA$) with the baseline. From the side panel, he marks prompts the previous model had struggled with using stars and examines them by selecting ``Starred Prompts''. Further, he applies a filter of ``Natural'' prompts to focus on his targeted topics, which he can toggle by choosing ``Filtered Prompts''. 

\textbf{G2. Annotate Issues (Fig.~\ref{fig:teaser}b).} After obtaining a high-level impression of the model performance, he enters the \textit{Annotation Mode} (Fig.~\ref{fig:annotation}) to conduct a detailed evaluation of model outputs. He defines three key issues he observed, including technical, artistic, and diverse issues, for which he selects the corresponding icons for each row of outputs to quickly annotate all prompts. For example, he marked a technical issue for $Baseline$ output of \textit{``A teddy bear skating in Times Square''} with a specific comment \textit{``distorted object''}. Meanwhile, he can share the annotation with collaborators to collectively annotate the results.

\textbf{G3. Analyze Performance (Fig.~\ref{fig:teaser}c).} Upon annotating issues for all prompts, Ben immediately sees the aggregate results of failure cases under the \textit{Analyze Page} (Fig.~\ref{fig:analyze}). As shown in Fig.~\ref{fig:teaser}c, he compares the issue count of the two models using summary bar charts, where he sees $ModelA$ (red) has fewer issues overall and improves in ``diverse'' issues significantly (left chart). A similar trend was found for the starred prompts (middle chart) and filtered prompts of ``Natural'' category (right chart). Meanwhile, he observed that $ModelA$ has zero diverse issues, significantly improving from the $Baseline$. Interestingly, there still seem to be several technical issues for $ModelA$.

\textbf{G4. Derive Insights (Fig.~\ref{fig:teaser}d).} Ben examines the individual model summary cards below the summary bar charts. He sees the full list of prompts and annotations for $ModelA$, and a collection of tags showing the comments he made (Fig.~\ref{fig:analyze}). As shown in Fig.~\ref{fig:teaser}d, he observes 6 occurrence of \textit{``distorted object''} and 4  \textit{``blurry''}. Combining the observations from summary bar charts, he concludes his evaluation that \textit{``ModelA has improved on diversity of motions but has introduced technical issues with object distortion and blurriness.''}
Finally, he shared his analysis with the team with an exported pdf.

\subsection{Interface Design}
\label{sec:interface}
We described key components of \genlens{} to support effective GenAI model output evaluation workflow. In a text-to-video task, each model outputs several \textit{instances} for each \textit{prompt}, which we refer to as an \textit{output}. Each prompt can have multiple outputs from different models.

\subsubsection{Discover Page}
As shown in Fig.~\ref{fig:teaser},
\textit{Discover Page} presents all model outputs with flexible filters and viewing options to support data overview (T1).
\textbf{Side Panel.} The user can upload the prompts and the corresponding image/video outputs from multiple models by providing the metadata and links to the model outputs.
To support comparison of model outputs, a user can choose which model(s) to show from the list of models the user imported. Further, the user can filter prompts based on bookmarks (star), filtering criteria, or the annotated issues, which can be dynamically updated.
This allow developers to flexibly assign prompt groups and quickly alter between categories without visual clutter. 
\textbf{Viewing Options.}
A user can toggle between a list or a gallery view to quickly skim through all outputs or carefully look at individual output. This requirement was based on the user feedback that high-level and detailed examination of outputs were both necessary.

\subsubsection{Annotation Mode}
As shown in Fig.~\ref{fig:annotation}, the user can click on the edit icon on the bottom left of \genlens{} interface to enter \textit{Annotation Mode}, which supports the user to annotate issues at the prompt level  (T2). The user can define their issue tags and icons, and go through the prompts to quickly annotate each output by selecting the icons. The user can also leave open comments for each output. The annotation can be save and shared in a JSON format, supporting collaborative model evaluation between developers.

\subsubsection{Analyze Page}
As shown in Fig.~\ref{fig:analyze}, \textit{Analyze Page} provides users with an annotation summary to analyze model performance and derive insights (T3). On the top, summary bar charts visualize the number of annotated issues across three categories, all prompts, starred prompts, and filtered prompts. This allows users to compare performance across models and categories. At the bottom, model summary cards aggregates the comments and annotations for each model. The user can quickly overview the issues and record their insights for the model in an open text field. Finally, they can export the summary page in a pdf to share across the team.

\subsection{Implementation}
\genlens{} is a web application with React-based front-end. The outputs are loaded to the interface at run time from URLs pointing to folders containing model output files (images/videos). \genlens{} uses localStorage API to save the annotations locally and persist them across page refreshes. Metadata are stored and loaded as JSON format, including prompt details and categories, model descriptions, and annotations.

\subsection{Design Iteration}
\genlens{} has gone through a two-stage iterative design process with three model developers before implementation. We first built a set of visual mock-ups to evaluate and prioritize different components. 
Based on the first round of design feedback, we improved on the design and built an interactive prototype using AdobeXD to verify the user flow. Several critical insights were obtained to inform our design decisions.

For example, in the \textit{Discover Page}, we iterated on the filtering view based on user feedback. Initially, we categorized prompts based on concepts, such as materials, animals, and colors. While categorizing prompts is very useful, user feedback suggests that a combination of filtering criteria is necessary as some prompts can belong to multiple categories. To simplify the filtering hierarchy and avoiding duplicate prompts, we designed filters to support multiple-select and dynamic update, which the user can toggle easily from the side bar. 

Further, \textit{Annotation Mode} has gone through drastic design change. Original design involved annotation at instance-level (single image/video), but users suggested annotating at output-level is more useful. Not only because annotating individual examples can be tedious, evaluating GenAI outputs for model development also focus more on the overall behaviors, such as diversity. As each developer focuses on different aspects during the evaluation, we provided customized issue tags and icons allowing free-form comments to support personalized annotation.
Additionally, users emphasized the importance to compare results between multiple models, as \textit{``Annotation is subjective and only make sense when you compare to something else''}, which informs the inclusion of outputs from multiple models for comparison.

In the \textit{Analysis Page}, we originally planned to provide more detailed model performance analysis such as scatter plots to show correlation between metrics or the proportion of issues among categories. However, feedback suggested that developers usually focus on overall patterns to derive actionable insights towards improving models. Therefore, highlighting a few key metrics of interest with clear comparison against other models was considered sufficient. Our resulting design provides clear comparison on the type and amount of failure cases, with overview of all annotations.

%% file: docs/5_discussion.tex

\section{Results}
To evaluate the usefulness of \genlens{}, we conducted a user study and gathered feedback for improvement. Below, we discussed the user study results and its implications for GenAI model development.

\subsection{User Evaluation}
\textbf{Procedure.} 
We conducted a user study with four GenAI model developers who are actively developing video generative models. All participants worked as either experienced ML engineers or research scientists with PhDs degrees in computer science with 5 to 10 years of experiences in machine learning. One participant had given feedback during the design iteration. The rest have not seen \genlens{} nor participated in the formative study before. 
Two datasets containing real model outputs to be evaluated by the participants were used in the study, each with 16 and 24 prompts and outputs from two models, respectively.

The user study has three parts: the experimenter first introduced \genlens{}' features, then the user evaluated their model outputs using \genlens{}, and finally participants rated the Technology Acceptance Model (TAM) questions and provided their feedback in the post-study interview. The study took around 30 minutes each.

\begin{figure}[t]
  \centering
  \includegraphics[width=\linewidth]{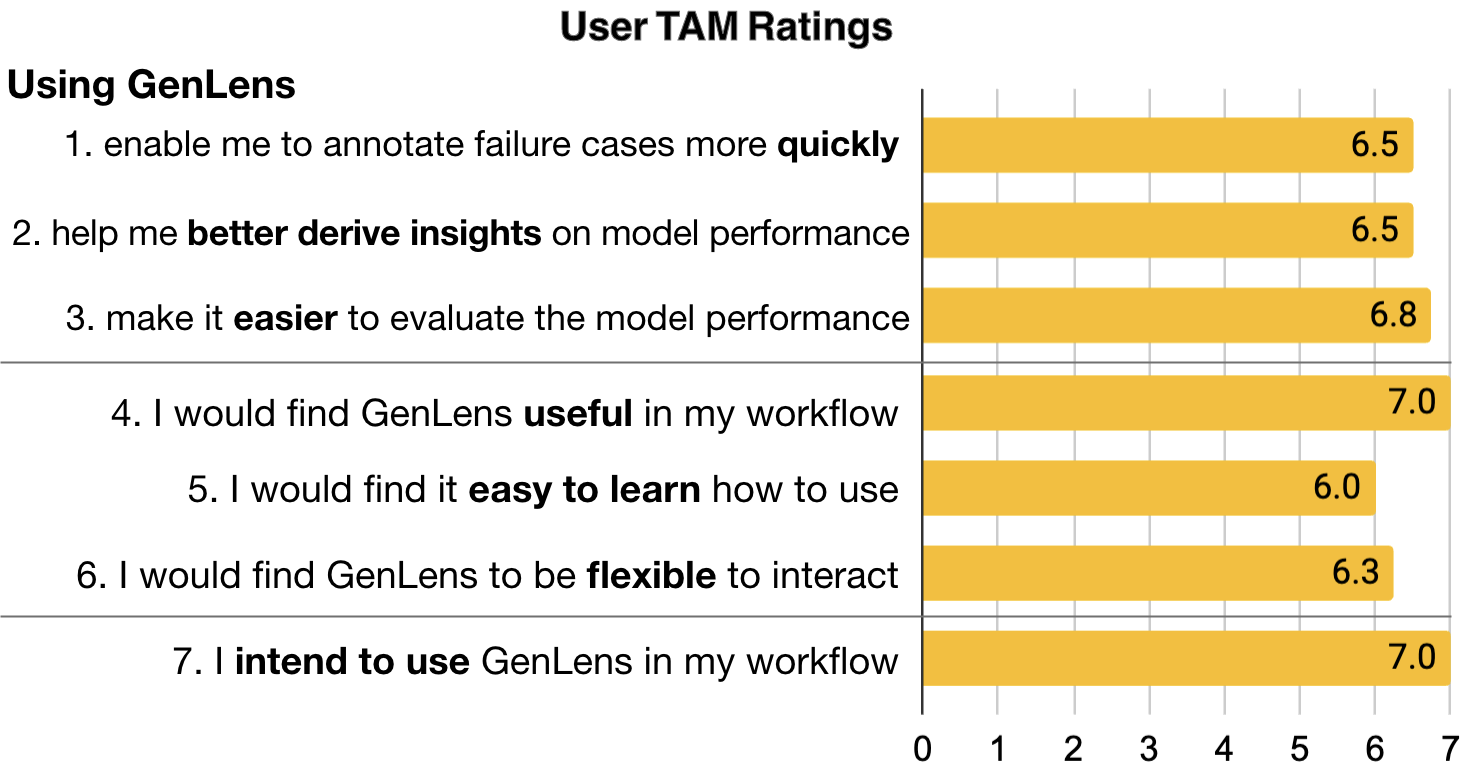}
  \caption{User TAM rating results. Participants rated \genlens{} as helpful (Q1-Q3), easy to use (Q4-Q6) to support their model evaluation. They also expressed high intent to use \genlens{} in their workflow (Q7).}
  \vspace{-4mm}
  \label{fig:rating}
\end{figure}

\textbf{Results.} Overall, as shown in Fig.~\ref{fig:rating}, participants expressed high level of satisfaction on perceived helpfulness (Q1-Q3), perceived ease of use (Q4-Q6) and intent to use (Q7) of \genlens{}. Particularly, they all found \genlens{} useful and intended to use it in their workflow, with a rating of 7 on a 7-point Likert scale. 

Participants were able to use \genlens{} to overview and annotate their model outputs during the user study with ease. 
One key observation was that users had very different focuses when evaluating the model outputs, and thus they required customized labels for the issues.
Some participants focused on comparison between two models, and used \textit{``better''}, \textit{``similar''}, \textit{``worse''} tags to directly quantify model performance. Others working on improving model quality focused more on different categories of technical issues, such as frame quality, motion, and distortion. 

Besides the user friendly annotation interface,
participants found \genlens{} very valuable in sharing their evaluation results and annotations, a critical part to facilitate more efficient communications between collaborators and support better insight verification. In addition, automatic aggregation of the annotations in bar charts and top comments help them effectively capture the issues to derive insights. They were able to immediately summarize the model issues based on the annotations, such as \textit{``there were many motion blur issues in this model.''} 

\textbf{Suggestions.} For improvement, participants requested more flexible way to view the individual instances, such as showing the original image/video aspects and allowing playing/pausing the video at different speeds for closer scrutiny. 
There were also suggestions regarding scalability, such as using pagination to allow loading outputs more effectively, with potential to extend to thousands of outputs. 
After the user study, these suggestions were implemented before deploying \genlens{} for internal use.
Further feedback needs to be collected to evaluate the usefulness of \genlens{} in the wild.  

\subsection{Implications for GenAI Model Development}

Two implications were derived from our design study of \genlens{}.

\textbf{Enhance collaboration in early-stage model evaluation.}
In developing large-scale GenAI models, various roles are involved at different stages, ranging from data collection, model design to the evaluation of results. 
To ensure the model quality conforms to human values, it is crucial to foster effective collaboration, particularly in the initial stages of model development. 
Our first-hand observations pointed to a significant gap in early-stage model evaluation, often due to the rush in AI development competition. 
Tools that enable collaborative and efficient performance evaluation are vital to prevent the overload of responsibility on a single developer and ensure a more thorough and balanced assessment of the model. This inclusion in the early-stage model development is crucial to enhance both the process and the quality of the ultimate GenAI models.

\textbf{Human-Centered GenAI Development.} Research in disciplines like HCI and Visualizations have been pivotal in devising new methods and techniques for innovative visualization and interaction, aimed at promoting human-centered technology development. 
However, the fast-paced AI field still faces significant challenges in ensuring effective human engagement, as highlighted by Chen et al.~\cite{chen2023next}. 
Our collaboration with AI developers indicates a notable gap in supporting essential visualization tasks during model evaluation, such as data comparison, annotation, and failure analysis.
While visualization researchers are often keen on tackling novel challenges, there is a pressing need to focus on developing simple yet effective solutions for these critical areas to ensure human-centric development in technology. 
Utilizing established human-centered research methods to tackle problems in emerging sectors like GenAI is essential for ensuring a balanced progression of technology. This study thus highlights and advocates for the importance of interdisciplinary work to develop practical, human-centric solutions in line with the rapid progression of AI technologies.

\subsection{Limitations \& Generalizability}
Although our formative study involving six GenAI model developers indicates a common workflow in generative technology development, it may not encompass the experiences of developers in other environments. Nevertheless, we anticipate that the challenges identified in the evaluation process are likely similar across different settings, as the GenAI community not only shares a common workflow but also strives to rapidly advance and refine their models in response to the fast-paced nature of the field.
Our tool \genlens{} was specifically tested with video outputs for text-to-video tasks. While its design elements are adaptable to other visual data and tasks, performance optimization is necessary, especially when handling and rendering large-scale datasets.
In addition,  while \genlens{}  was designed to address early-stage model output evaluations for developers, we also see the potential for adoption in post-deployment assessments of model outputs targeted at end users.

%% file: docs/6_conclusion.tex

\section{Conclusions and Future Work}
This paper introduces a design study focused on the systematic assessment of visual outputs from generative AI (GenAI) models, particularly those generating images and videos.
Our formative study with GenAI model developers identified significant gaps in early-stage model evaluation. To address this, we developed \genlens{}, a visual analytic interface that streamlined the annotation and analysis of failure cases to facilitate a systematic evaluation of the model outputs. A user study with active GenAI model developers demonstrated the usefulness of \genlens{}. Our study results also pointed to needs for tools that enhance effective human-in-the-loop involvement in model development.

For future work, \genlens{} can be enhanced with algorithms for automatically identify and categorize failure cases, reducing the manual workload for developers. Additionally, extending \genlens{} to work with a wider range of visual GenAI models and tasks will be feasible, such as 3D data.
We also plan to collect direct feedback on the real-world impact of such evaluation on model development from deploying \genlens{} in actual work environments.

%% file: docs/acknowledgement.tex
\acknowledgments{\scriptsize
This work is supported by NSF grant III-2107328 and Adobe Research.
We would like to express our gratitude to 
Ali Aminian,
Zoya Bylinskii, 
Fabian Caba Heilbron,
Tobias Hinz,
Kushal Kafle,
Mingi Kwon,
Difan Liu,
Simon Niklaus, 
Esme Xu, 
and Yang Zhou for their feedback.
We thank the reviewers for their valuable comments.
}